\documentclass[12pt]{article}

\usepackage[textheight=22.5truecm,textwidth=16.8truecm,top=2.95truecm,left=2.15truecm]{geometry}
\usepackage{
amsmath,
amsthm,
amssymb,
ascmac,
bm,
tikz,
braket,
mathrsfs,
graphicx,
color,
hyperref,
cite,
titlesec,
comment
}
\allowdisplaybreaks

\hypersetup{colorlinks,bookmarksopen,bookmarksnumbered,citecolor=black,linkcolor=black,linktocpage,pdfstartview=FitH,urlcolor=black}

\numberwithin{equation}{section}

{\makeatletter
\long\def\@makefntext#1{\parindent 1em\noindent 
\@hangfrom{\hbox to 1.8em{\hss$^{\@thefnmark}$}}#1}
\makeatother}

\def\fnum@figure{\textbf{\figurename\nobreakspace\thefigure}}
\def\fnum@table{\textbf{\tablename\nobreakspace\thetable}}
\long\def\@makecaption#1#2{%
  \vskip\abovecaptionskip
  \sbox\@tempboxa{\small #1. #2}%
  \ifdim \wd\@tempboxa >\hsize
    \small #1. #2\par
  \else
    \global \@minipagefalse
    \hb@xt@\hsize{\hfil\box\@tempboxa\hfil}%
  \fi
  \vskip\belowcaptionskip}

\renewcommand{\l}[0]{\left}
\renewcommand{\r}[0]{\right}
\renewcommand{\d}[0]{\mathrm{d}}

\newcommand{\lcmt}[2]{\l[#1,#2\r]}

\newcommand{\cmt}[2]{[#1,#2]}
\newcommand{\acmt}[2]{\{#1,#2\}}

\newcommand{\BRST}[0]{Q_\mathrm{B}}
\newcommand{\tv}[0]{\mathrm{tv}}
\newcommand{\tvop}[1]{U#1 U^{-1}}
\newcommand{\interior}[1]{\mathcal I_{#1}}
\newcommand{\Lie}[1]{\pounds_{#1}}
\newcommand{\tLie}[1]{\tilde\pounds_{#1}}
\newcommand{\tinterior}[1]{\tilde{\mathcal I}_{#1}}

\title{\hfill\parbox{3cm}{\normalsize KUNS-2927}\\[12pt]
Generating string field theory solutions \\
with matter operators from \textit{KBc} algebra
}

\author{Hiroyuki Hata\footnote{Professor emeritus of Kyoto University,
hata.hiroyuki.3@gmail.com},\quad
Daichi Takeda,$^1$\footnote{takedai@gauge.scphys.kyoto-u.ac.jp}\quad
Jojiro Yoshinaka$^1$\footnote{george.yoshinaka@gauge.scphys.kyoto-u.ac.jp}\\[12pt]
\textit{$^1$Department of Physics, Kyoto University, Kyoto 606-8502, Japan}
}

\date{}

\begin{document}
\maketitle
\begin{abstract}
	The \textit{KBc} algebra is a subalgebra that has been used to construct classical solutions in Witten's open string field theory, such as the tachyon vacuum solution.
	The main purpose of this paper is to give various operator sets that satisfy the \textit{KBc} algebra.
	In addition, since those sets can contain matter operators arbitrarily, we can reproduce the solution of Kiermaier, Okawa and Soler, and that of Erler and Maccaferri.
	Starting with a single D-brane solution on the tachyon vacuum, we replace the original \textit{KBc} in it with an appropriate set to generate each of the above solutions.
	Thus, it is expected that the \textit{KBc} algebra, combined with the single D-brane solution, leads to a more unified description of classical solutions.
\end{abstract}

\newpage

\tableofcontents

\section{Introduction}
A non-perturbative string theory should have an ability to describe the dynamics of the background, and should be formulated in a background independent way.
String field theory (SFT) is a second-quantized string theory, where backgrounds are determined as classical solutions of the EOM.
Since we can switch between backgrounds by shifting the string field, the background independence is also assured.
Witten's bosonic open string field theory \cite{Witten:1985cc} is known for that its EOM is relatively easy to analyze classically.

In Witten's SFT, the background is defined through a first-quantized string theory which is described by a world-sheet CFT, and the string field $\Psi$ is a composite operator of the reference CFT.
On the other hand, other backgrounds are realized by classical field configurations,  solutions of the EOM, $\BRST \Psi + \Psi^2=0$, where $\BRST$ is the BRST operator of the reference CFT.

Since a bosonic string theory is unstable, any D-brane system will decay to the tachyon vacuum \cite{Sen:1998sm,Sen:1999mg},  regardless of what CFT we choose at the beginning.
Thus, this condensation phenomena must be characterized without any specific information of the reference CFT.
In fact, the tachyon vacuum solution found in \cite{Schnabl:2005gv,Erler:2009uj} can be described by the so-called \textit{KBc algebra} \cite{Okawa:2006vm}, which is a universal subalgebra in the sense that any CFT possesses it.
This universality of the \textit{KBc} algebra guarantees that any D-brane system falls into the tachyon vacuum.

However, in order to establish the background independence explicitly, any background should be expressed by a solution of the EOM on the reference D-brane system.
Since a background is determined by a world-sheet CFT, the problem we would like to solve is whether we can find a solution that corresponds to a given CFT.
In particular, if we consider a one-parameter family of CFTs which includes the reference CFT, can we interpret it as a one-parameter family of SFT solutions?
This problem has been addressed not only with the \textit{KBc} algebra, but also with matter operators of the reference CFT.

One type of such one-parameter modifications is the \textit{marginal deformation}, which keeps the conformal invariance of the reference CFT.
The marginal deformation means adding to the CFT a boundary term which is characterized by a matter primary operator having weight 1.
There have been many attempts to find solutions which correspond to marginally deformed CFTs \cite{Kiermaier:2007ba,Schnabl:2007az,Fuchs:2007yy,Kiermaier:2007vu,Erler:2007rh,Okawa:2007ri,Okawa:2007it,Fuchs:2007gw,Kiermaier:2007ki,Kiermaier:2010cf}.
Among those attempts, the solution of Kiermaier, Okawa, and Soler (KOS) \cite{Kiermaier:2010cf} is known as a successful analytical solution, where the {\em boundary condition changing operators} (bcc operators) were introduced.
Since marginal deformations add boundary terms, their effects are regarded as changes of boundary conditions; the bcc operators play the role to change a boundary condition to another.
The KOS solution is described by the bcc operators and the \textit{KBc} algebra.
Though valid only when the matter operators are regular,\footnote{
Here ``regular" means having finite self-OPE.}
it was generalized to non-regular cases by Erler and Maccaferri \cite{Erler:2014eqa, Erler:2019fye}.
However, some questions like how to explicitly find unknown bcc operators are still left, even though their approach has advanced our understanding of the background structure.

In this paper, we revisit the \textit{KBc} algebra to deepen our understanding further.
In \cite{Hata:2021lqz}, we found operator sets which consist of the original $K,B$ and $c$ and satisfy the algebraic relations of the \textit{KBc} algebra.
We call these operator sets representations of the \textit{KBc} algebra.
Extending the previous method, we find in this paper a larger family of representations, which can arbitrarily contain matter operators.
In addition, we can map the representations to the \textit{KBc} algebra on the tachyon vacuum.
By using this, we explicitly show that each of the KOS and the Erler-Maccaferri solutions is reproduced by a specific representation of the \textit{KBc} algebra.
Although our work is mostly devoted to finding what kind of representation gives each solution, we will also discuss a potential ability of our framework to connect the \textit{KBc} algebras in different CFTs.
This is expected to lead to a deeper understanding of the whole background structure.

This paper is organized as follows.
In section \ref{sec: representation}, we introduce various representations of the \textit{KBc} algebra with matter operators, both on a single D-brane solution and on the tachyon vacuum.
Then in section \ref{sec: solutions}, we apply the latter to construct the two solutions, the KOS and the Erler-Maccaferri solutions.
In section \ref{sec: discussions}, we summarize the paper and discuss how to find unknown physically meaningful solutions by using our framework.
In appendices \ref{app: KBc construction} and \ref{app: EM detail}, we present the technical details used in the text.
We discuss representations of the extended algebra including matter operators in appendix \ref{app: KBc manifold}.

\section{Representations of the \textit{KBc} algebra}\label{sec: representation}
In this section, we introduce representations of the \textit{KBc} algebra containing arbitrary number of matter operators.
We first give representations on a single D-brane, and next on the tachyon vacuum.
The latter representations will be used later in section \ref{sec: solutions}, and hence play a key role in this paper.
For other attempts to consider representations of the \textit{KBc} algebra, see  \cite{Erler:2010zza,Masuda:2012kt,Erler:2012dz,Hata:2011ke,Mertes:2016vos}.

\subsection{Representations on a single D-brane}\label{subsec: KBc on D-brane}
Here we consider the world-sheet CFT which describes a single D-brane.
As is well known, the \textit{KBc} algebra is a subalgebra consisting of three operators $K,B$ and $c$ in the CFT, which satisfy the following relations \cite{Okawa:2006vm}:
\begin{align}
	&\cmt{K}{B}=0,\qquad \acmt{B}{c}=1,\qquad B^2=0,\qquad c^2=0,\label{eq: commutators}\\
	&\qquad \BRST K = 0,\qquad \BRST B = K,\qquad \BRST c = cKc.\label{eq: BRST operations}
\end{align}
Originally, $K,B$ and $c$ appearing in the above relations are defined in the sliver frame as
\begin{align}
	K = \int_{-i\infty}^{i\infty} \frac{\d z}{2\pi}\,T(z),\qquad
	B = \int_{-i\infty}^{i\infty}\frac{\d z}{2\pi}\,b(z),\qquad
	c = c(0).\label{eq: fundamental}
\end{align}
We call the triad $(K,B,c)$ in \eqref{eq: fundamental} the \textit{basic representation} in this paper, and $(K,B,c)$ always denotes the basic representation.
In the following, we will find other triads which satisfy \eqref{eq: commutators} and \eqref{eq: BRST operations} with the same BRST operator $\BRST$ on a single D-brane.
We call them representation of the \textit{KBc} algebra (on a single D-brane) or \textit{KBc}-representation for short.

Extending the idea of \cite{Hata:2021lqz}, we find that the following triad $(K(\xi),B(\xi),c(\xi))$ is also a representation of the \textit{KBc} algebra:
\begin{align}
	K(\xi) = \BRST \xi^1,\qquad
	B(\xi) = \xi^1,\qquad
	c(\xi) = e^{-i\BRST \xi^2}c e^{iK\Xi^2}\frac{B}{\Xi^1}e^{-i\Xi^2 K}c e^{i\BRST \xi^2}.
	\label{eq: representation xi}
\end{align}
Let us explain the ingredients  appearing in this expression.
First,  $\xi = (\xi^1,\xi^2)$ carries ghost number $-1$ and consists
of $(K,B,c)$ and matter operators.\footnote{
When matter operators are not included,
$\xi^1$ and $\BRST\xi^2$ in this paper correspond to $B e^{\xi^1}$ and
$\xi^2$ in \cite{Hata:2021lqz} (see (3.12) there), respectively.
}
Since matter operators commute with $B$ and $c$, $\xi$ can be expressed as
\begin{align}
	\xi = (\xi^1,\xi^2) = B(\Xi^1,\Xi^2) = (\Xi^1,\Xi^2)B,
\end{align}
with $\Xi^{1,2}$ consisting of $K$ and matter operators.
Note that $K\Xi^2 = \Xi^2 K$ does not hold in general, and that the triad \eqref{eq: representation xi} is real if $\Xi^{1,2}$ is real.\footnote{Real means that the quantity is self double conjugate.}
Of course, all \textit{KBc}-representations cannot be expressed as \eqref{eq: representation xi}.
A more general form of \textit{KBc}-representations is given in appendix \ref{app: KBc construction}.

Leaving the construction of \eqref{eq: representation xi} in appendix \ref{app: KBc construction}, here let us just confirm that \eqref{eq: representation xi} satisfies \eqref{eq: commutators} and \eqref{eq: BRST operations}.
Nontrivial relations are $\{B(\xi),c(\xi)\} = 1$ and $\BRST c(\xi) = c(\xi)K(\xi)c(\xi)$.
To show the former, we first note that
\begin{align}
	B \BRST\xi^{1,2} &= B(K\Xi^{1,2} - B\BRST \Xi^{1,2}) = BK\Xi^{1,2} = K\Xi^{1,2} B,\nonumber\\ 
	(\BRST \xi^{1,2}) B &= \Xi^{1,2}KB = B\Xi^{1,2}K.\label{eq: B and xi}
\end{align}
Then we find that
\begin{align}
	B(\xi)c(\xi) 
	&= \Xi^1 B e^{-i\BRST\xi^2}c e^{iK\Xi^2}\frac{B}{\Xi^1}e^{-i\Xi^2 K}c e^{i\BRST \xi^2} 
	= \Xi^1 e^{-iK\Xi^2}BcB e^{iK\Xi^2}\frac{1}{\Xi^1}e^{-i\Xi^2 K}c e^{i\BRST \xi^2}\nonumber\\
	&=\Xi^1 e^{-iK\Xi^2}Be^{iK\Xi^2}\frac{1}{\Xi^1}e^{-i\Xi^2 K}c e^{i\BRST \xi^2}
	=e^{-i\Xi^2 K}Bc e^{i\BRST \xi^2}=e^{-i\BRST \xi^2}Bc e^{i\BRST \xi^2},
\end{align}
and similarly,
\begin{align}
	c(\xi)B(\xi) = e^{-i\BRST \xi^2}cB e^{i\BRST \xi^2}.\label{eq: cB}
\end{align}
Therefore $\{B(\xi),c(\xi)\} = 1$ holds due to $\acmt{B}{c}=1$.

Next, for showing $\BRST c(\xi) = c(\xi)K(\xi)c(\xi)$, we rewrite the r.h.s.\ as
\begin{align}
	c(\xi)K(\xi)c(\xi) &= -\BRST (c(\xi)B(\xi)c(\xi)) + (\BRST c(\xi))B(\xi)c(\xi) + c(\xi)B(\xi)\BRST c(\xi).\nonumber\\
	 &=\BRST c(\xi) - (\BRST c(\xi))c(\xi)B(\xi) - B(\xi)c(\xi) \BRST c(\xi),
\end{align}
where we have used $\{B(\xi),c(\xi)\} = 1$, which we have just proven, and in particular that $c(\xi)B(\xi)c(\xi) = c(\xi)$.
Since we can directly confirm from \eqref{eq: representation xi} that $\BRST c(\xi)$ is written as
\begin{align}
	\BRST c(\xi) = e^{-i\BRST\xi^2}c\bigl[\quad\cdots\quad\bigr]ce^{i\BRST\xi^2},
\end{align}
we see that $(\BRST c(\xi))c(\xi) = c(\xi)\BRST c(\xi) = 0$, and hence that $\BRST c(\xi) = c(\xi)K(\xi)c(\xi)$ holds.

Note that \eqref{eq: representation xi} does not in general respect relations which are not in \eqref{eq: commutators} and \eqref{eq: BRST operations} but are satisfied by the basic representation; for example, $cKcKc = 0$ which follows from $[c,K] = \partial c$.
An example is $\xi = B(1+\epsilon K,0)$ with $\epsilon$ infinitesimal and real.
For this $\xi$, one can show $c(\xi)K(\xi)c(\xi)K(\xi)c(\xi) = \epsilon cKcK^2 c + O(\epsilon^2) \neq 0$.

\subsection{Representations on the tachyon vacuum}\label{subsec: tv rep}
In the previous subsection, we introduced \textit{KBc}-representations \eqref{eq: representation xi} on a D-brane.
As we know that the tachyon vacuum is the most basic background, it seems useful to find various representations of the \textit{KBc} algebra in its corresponding CFT with a trivial BRST cohomology.
In this case, $\BRST$ in the \textit{KBc} algebra is replaced by $Q_\tv := \BRST + [\Psi_0,\cdot \}$,\footnote{$[A,B\} := AB - (-1)^{|A||B|}BA$ with $|A|:=+1$ ($-1$) if $A$ is Grassmann-even (-odd).}
where $\Psi_0$ is the tachyon vacuum solution; we call this algebra \textit{KBc}$_\tv$ algebra.

We adopt the simple solution \cite{Erler:2009uj} as the tachyon vacuum solution:
\begin{align}
	&\Psi_0 = \tvop{\BRST} = \frac{1}{\sqrt{1+K}}c(1+K)Bc\frac{1}{\sqrt{1+K}},\label{eq: tachyonj vacuum solution}\\
	&U = \l(\sqrt{1+K}-\frac{1}{\sqrt{1+K}}Bc\r)\frac{1}{\sqrt{K}},\label{eq: U}\\
	&U^\ddag = U^{-1} = \l(\sqrt{K} + \frac{1}{\sqrt{K}}Bc\r)\frac{1}{\sqrt{1+K}}\label{eq: U inverse}.
\end{align}
The BRST operator $Q_\tv$ of the \textit{KBc}$_\tv$ algebra is defined by this $\Psi_0$.
Since we have
\begin{align}
	(\tvop{X})(\tvop{Y}) = \tvop{XY},\qquad Q_\tv (\tvop{X}) = \tvop{(\BRST X)},
	\label{eq: property of tv}
\end{align}
for any $X$ and $Y$, sandwiching \eqref{eq: representation xi} between $U$ and $U^{-1}$ directly gives \textit{KBc}$_\tv$-representations,
\begin{align}
	&K(\xi)_\tv = \tvop{(\BRST \xi^1)} ,\qquad
	B(\xi)_\tv = \tvop{\xi^1} ,\qquad\nonumber\\
	&c(\xi)_\tv = \tvop{e^{-i\BRST \xi^2}c e^{iK\Xi^2}\frac{B}{\Xi^1}e^{-i\Xi^2 K}c e^{i\BRST \xi^2}}.
	\label{eq: tv representation xi}
\end{align}
Namely, this triad satisfies
\begin{align}
	&\cmt{K(\xi)_\tv}{B(\xi)_\tv }=0,\qquad \acmt{B(\xi)_\tv }{c(\xi)_\tv }=1,\qquad B(\xi)_\tv ^2=0,\qquad c(\xi)_\tv ^2=0,\\
	&\qquad Q_\tv  K(\xi)_\tv  = 0,\qquad Q_\tv B(\xi)_\tv  = K(\xi)_\tv,\qquad Q_\tv  c(\xi)_\tv  = c(\xi)_\tv K(\xi)_\tv c(\xi)_\tv.
\end{align}
In the following, we define $X_\tv $ for any $X$ as\footnote{
In this notation, \eqref{eq: property of tv} is written as
$X_\tv Y_\tv  = (XY)_\tv$ and $Q_\tv X_\tv  = (\BRST X)_\tv $.
}
\begin{align}
	X_\tv := U X U^{-1}.\label{eq: subscript tv}
\end{align}
Using this notation and \eqref{eq: property of tv}, we can rewrite \eqref{eq: tv representation xi} as
\begin{align}
	&K(\xi)_\tv = Q_\tv\xi^1_\tv ,\quad
	B(\xi)_\tv = \xi^1_\tv ,\quad
	c(\xi)_\tv = e^{-iQ_\tv \xi^2_\tv}c_\tv e^{iK_\tv\Xi^2_\tv}\frac{B_\tv}{\Xi^1_\tv}e^{-i\Xi^2_\tv K_\tv}c_\tv e^{iQ_\tv \xi^2_\tv}.
	\label{eq: tv representation xi 2}
\end{align}

\section{Construction of classical solutions}\label{sec: solutions}
In this section, we explicitly show that \textit{KBc}$_\tv$-representations reproduce the KOS and the Erler-Maccaferri solutions.
Hereafter, the string field $\Psi$ represents the deviation from the tachyon vacuum, meaning that its EOM is given as
\begin{align}
	Q_\tv \Psi + \Psi^2 = 0.\label{eq: EOM}
\end{align}
In \cite{Erler:2014eqa}, the single D-brane solution on the tachyon vacuum, $\Psi = -\Psi_0$, is deformed by using bcc operators \cite{Kiermaier:2010cf} as $-\Sigma\Psi_0\bar\Sigma$, which was shown to give solutions corresponding to various backgrounds.
Our strategy here is to deform $-\Psi_0$ by using \textit{KBc}$_\tv$-representations to express backgrounds.

\subsection{Generating other solutions from a single D-brane solution}
We first rewrite $-\Psi_0$ in terms of $(K_\tv,B_\tv,c_\tv)$, which is obtained by applying \eqref{eq: subscript tv} to the basic representation $(K,B,c)$.
Noting that
\begin{align}
	U_\tv = \tvop{U} = U,
\end{align}
we can rewrite $-\Psi_0$ as
\begin{align}
	- \Psi_0 = -\tvop{\BRST} &= U^{-1}U(\BRST U)U^{-1} = U^{-1}Q_\tv(UUU^{-1})= U^{-1}_\tv Q_\tv U_\tv,
\end{align}
where we have also used \eqref{eq: property of tv} at the third equality.
By using \eqref{eq: property of tv} again, we see from \eqref{eq: U} and \eqref{eq: U inverse} that
\begin{align}
	U &= U_\tv = \l(\sqrt{1+K_\tv}-\frac{1}{\sqrt{1+K_\tv }}B_\tv c_\tv \r)\frac{1}{\sqrt{K_\tv }},\label{eq: tv U} \\
	U^{-1} &=U^{-1}_\tv = \l(\sqrt{K_\tv } + \frac{1}{\sqrt{K_\tv }}B_\tv c_\tv \r)\frac{1}{\sqrt{1+K_\tv }},\label{eq: tv U inverse}
\end{align}
and hence we can regard $-\Psi_0$ as a string field expressed completely by $(K_\tv ,B_\tv ,c_\tv )$:
\begin{align}
	-\Psi_0 =U^{-1}_\tv Q_\tv U_\tv  =: \Psi_1(K_\tv ,B_\tv ,c_\tv ).\label{eq: D-brane}
\end{align}

Since we only need \textit{KBc}$_\tv$ algebra to show that \eqref{eq: D-brane} is a solution to the EOM \eqref{eq: EOM}, 
string field $\Psi_1(K(\xi)_\tv,B(\xi)_\tv,c(\xi)_\tv)$ is also a solution.
Explicitly, we have
\begin{align}
	&\Psi_1(K(\xi)_\tv,B(\xi)_\tv,c(\xi)_\tv )= [U_\tv ^{-1}Q_\tv U_\tv]_\xi
	 = -\l[\frac{1}{\sqrt{K_\tv }}c_\tv B_\tv \frac{K_\tv^2}{1+K_\tv }c_\tv \frac{1}{\sqrt{K_\tv }}\r]_\xi,
	 \label{eq: tachyon Psi_1}
\end{align}
where $[\cdots]_\xi$ is the symbol to mean that $(K_\tv,B_\tv,c_\tv)$ inside the bracket is replaced by $(K(\xi)_\tv,B(\xi)_\tv,c(\xi)_\tv )$.
If we choose $\xi$ real, the triad $(K(\xi)_\tv,B(\xi)_\tv,c(\xi)_\tv )$ is real, and hence the solution \eqref{eq: tachyon Psi_1} is.
In the rest of this section, we will show that \eqref{eq: tachyon Psi_1} reproduces the KOS and the Erler-Maccaferri solutions by choosing suitable real $\xi$'s.

\subsection{Reproducing KOS solution for marginal deformation}\label{subsec: KOS}
The KOS solution \cite{Kiermaier:2010cf} describes a marginal deformation of the CFT characterized by a regular matter primary operator $V$ with its weight $1$, where ``regular" means that $cV$ has a vanishing self-OPE.
They introduced the bcc operators $\sigma$ and $\bar\sigma$, which are matter primary operators of weight 0 and are related to $V$ through
\begin{align}
	V = -\sigma\partial \bar\sigma,\qquad
	e^{-\alpha(K+V)} = \sigma e^{-\alpha K} \bar\sigma.\label{eq: sigma and V}
\end{align}
The following is also an important relation:
\begin{align}
	\sigma\bar\sigma = \bar\sigma \sigma = 1.\label{eq: sigmasigma}
\end{align}
The solution is given as
\begin{align}
	\Psi_\mathrm{KOS} = -\frac{1}{\sqrt{1+K}}c(1+K)\sigma \frac{B}{1+K}\bar\sigma(1+K)c\frac{1}{\sqrt{1+K}},\label{eq: KOS}
\end{align}
which satisfies \eqref{eq: EOM}.

Our problem is to find $\xi$ such that
\begin{align}
	[\Psi_1(K_\tv,B_\tv,c_\tv)]_\xi = \Psi_\mathrm{KOS}.
	\label{eq: original KOS condition}
\end{align}
Multiplied by $U^{-1}$ and $U$, this condition reads
\begin{align}
	\frac{1}{\sqrt{K(\xi)}}c(\xi)\frac{K(\xi)^2}{1 + K(\xi)}B(\xi)c(\xi)\frac{1}{\sqrt{K(\xi)}}  = 
	\frac{1}{\sqrt{K}}cK\sigma\frac{B}{1+K}\bar\sigma Kc\frac{1}{\sqrt{K}}~  (=U^{-1}\Psi_\mathrm{KOS}U),
	\label{eq: marginal condition}
\end{align}
where we have used \eqref{eq: tachyon Psi_1} and that $U^{-1}[X_\tv]_\xi U$ is equal to $X$ with $(K,B,c)$ in it replaced by $(K(\xi),B(\xi),c(\xi))$.

By considering $B\eqref{eq: marginal condition}B$, we obtain a necessary condition for $\Xi^1$:
\begin{align}
	B\frac{1}{\sqrt{K\Xi^1}}\frac{1}{\Xi^1}\frac{(\Xi^1K)^2}{1+\Xi^1 K}\frac{1}{\sqrt{\Xi^1 K}}
	= B\sqrt{K}\sigma \frac{1}{1+K}\bar\sigma\sqrt{K}.\label{eq: B and B}
\end{align}
In deriving the l.h.s., we have used \eqref{eq: B and xi} and $B = B(\xi)/\Xi^1$.
Since $B$ commutes with all the other operators appearing in \eqref{eq: B and B}, we can remove $B$ from it.
Using $\sqrt{K\Xi^1} = K\sqrt{\Xi^1 K}K^{-1}$, \eqref{eq: B and B} with $B$ removed is rewritten as
\begin{align}
	K\frac{1}{1 + \Xi^1 K} = \sqrt{K}\sigma\frac{1 }{1 + K}\bar\sigma\sqrt{K},\label{eq: condition on Xi^1}
\end{align}
which determines $\Xi^1$:
\begin{align}
	\Xi^1 = 1+\frac{1}{\sqrt{K}}\l[\l(\sigma\frac{1}{1+K}\bar\sigma \r)^{-1} -1-K\r]\frac{1}{\sqrt{K}} = 1 + \frac{1}{\sqrt{K}}V\frac{1}{\sqrt{K}}.\label{eq: xi^1 marginal}
\end{align}
Here we have used \eqref{eq: sigmasigma} and $\sigma K \bar\sigma = K+V$, the latter of which follows from the $\alpha$-derivation of the second equation in \eqref{eq: sigma and V}.
This $\Xi^1$ is real since we are taking a real $V$ (hence
$\sigma^\ddag = \bar\sigma$).

Next, let us determine $\Xi^2$.
For this purpose, we multiply \eqref{eq: marginal condition} by $B$ only from the left side to get a necessary condition for $\Xi^2$:
\begin{align}
	\sqrt{K}\sigma \frac{1}{1+K}\bar\sigma\sqrt{K}\l[ \sqrt{\Xi^1K}e^{-i\Xi^2K}Bce^{i\BRST \xi^2}\frac{1}{\sqrt{\BRST\xi^1}}	-\sqrt{K}Bc\frac{1}{\sqrt{K}}\r]=0,
	\label{eq: sufficient condition of marginal}
\end{align}
where we have used \eqref{eq: condition on Xi^1}.
As proven in appendix \ref{app: EM detail}, \eqref{eq: marginal condition} actually follows from the condition that the quantity inside the bracket in \eqref{eq: sufficient condition of marginal} vanishes.
Thus our task is to determine $\Xi^2$ from
\begin{align}
\sqrt{\Xi^1K}e^{-i\Xi^2K}Bc\,e^{i\BRST \xi^2}\frac{1}{\sqrt{\BRST\xi^1}}
=\sqrt{K}Bc\frac{1}{\sqrt{K}}.
\label{eq: xi^2 condition}
\end{align}

For this purpose, note first the following expression for the quantity
on the l.h.s.\ of \eqref{eq: xi^2 condition}
obtained by using the singular homotopy operator $B/K$ and the formula
\eqref{eq: B and xi}: 
\begin{align}
e^{i\BRST \xi^2}\frac{1}{\sqrt{\BRST\xi^1}}
 &=\BRST\!\left(\frac{B}{K}e^{i\BRST \xi^2}\frac{1}{\sqrt{\BRST\xi^1}}\right)
 =\BRST\!\left(\frac{B}{K}e^{iK\Xi^2}\frac{1}{\sqrt{K\Xi^1}}\right)
 =\BRST\!\left(\frac{B}{\sqrt{K}}F^{-1}\frac{1}{K}\right)\label{eq:B/sqrtKG},
\end{align}
where we have defined $F$ by
\begin{align}
    F:= \sqrt{\Xi^1K}e^{-i\Xi^2K}\frac{1}{\sqrt{K}}.\qquad 
\label{eq: def of F and G}
\end{align}
Using \eqref{eq:B/sqrtKG}, the l.h.s.\ of \eqref{eq: xi^2 condition}
is rewritten as
\begin{align}
\sqrt{\Xi^1K}e^{-i\Xi^2K}Bc\,e^{i\BRST \xi^2}\frac{1}{\sqrt{\BRST\xi^1}}
&=F\sqrt{K}Bc\,
\BRST\!\left(\frac{B}{\sqrt{K}}F^{-1}\frac{1}{K}\right)\nonumber\\
    &=F\sqrt{K}Bc\sqrt{K}F^{-1}\frac{1}{K}+B(\BRST F)F^{-1}\frac{1}{K},\label{eq:find xi^2 with F}
\end{align}
where we have used
$\BRST F^{-1}=-F^{-1}(\BRST F)F^{-1}$ at the second equality.
Therefore, the condition \eqref{eq: xi^2 condition} for $\Xi^2$
(and hence for $F$) is now reduced to
\begin{align}
\BRST F=\cmt{\sqrt{K}c\sqrt{K}}{F}.\label{eq: BBRST F}
\end{align}

Besides \eqref{eq: BBRST F}, we have to take into account that $F$
is defined as a particular quantity \eqref{eq: def of F and G}.
Using $(\Xi^{1,2})^\ddag = \Xi^{1,2}$ (we will check
$(\Xi^2)^\ddag=\Xi^2$ below for the resultant $\Xi^2$), we obtain
\begin{align}
FF^\ddag &=\sqrt{\Xi^1 K}e^{-i \Xi^2 K}\frac{1}{K}e^{iK\Xi^2}\sqrt{K\Xi^1}
=\sqrt{\Xi^1 K}e^{-i \Xi^2 K}e^{i\Xi^2 K}\sqrt{\Xi^1 K}\frac{1}{K}
=\Xi^1 .
\label{eq:FF^ddag=Xi^1}
\end{align}
Let us take the simplest one as $F$ satisfying \eqref{eq:FF^ddag=Xi^1}:
\begin{equation}
F=\sqrt{\Xi^1} .
\label{eq:F=sqrtXi^1}
\end{equation}
This $F$ actually satisfies \eqref{eq: BBRST F} as we shall show.
From \eqref{eq: xi^1 marginal}, $\sqrt{\Xi^1}$ is Taylor-expanded as
\begin{align}
	\sqrt{\Xi^1} = \sum_{n=0}^\infty a_n\l(\frac{1}{\sqrt{K}}V\frac{1}{\sqrt{K}}\r)^n
		= \frac{1}{\sqrt{K}}\sum_{n=0}^\infty a_n\l(V\frac{1}{K}\r)^n\sqrt{K},\label{eq: a_n}
\end{align}
with numerical coefficients $a_n$.
 From the fact that $V$ is a matter primary operator of weight 1 and $\cmt{V}{c} = 0$, we have
 \begin{align}
	\BRST V = \cmt{K}{cV}\label{eq: BRST V},
\end{align}
from which we also have
\begin{align}
	\BRST	\l(V\frac{1}{K}\r)= \lcmt{Kc}{V\frac{1}{K}}.\label{eq: BRST V/K}
\end{align}
Since both $\BRST$ and the commutator follow the Leibniz rule, we have
\begin{align}
	\BRST	\l(V\frac{1}{K}\r)^n =  \lcmt{Kc}{\l(V\frac{1}{K}\r)^n},
\end{align}
for all $n \in \mathbb N$.
This implies that $F=\sqrt{\Xi^1}$ given by the series \eqref{eq: a_n}
satisfies \eqref{eq: BBRST F}.
From \eqref{eq: def of F and G},
$\Xi^2$ corresponding to $F$ of \eqref{eq:F=sqrtXi^1} is given in
terms of $\Xi^1$ by
\begin{equation}
e^{-i\Xi^2 K}=\frac{1}{\sqrt{\Xi^1 K}}\sqrt{\Xi^1}\sqrt{K},
\qquad
\Xi^2=i\ln\!\left(\frac{1}{\sqrt{\Xi^1 K}}\sqrt{\Xi^1}\sqrt{K}\right)
\frac{1}{K}.
\label{eq: xi^2 marginal}
\end{equation}

Finally, let us confirm $(\Xi^2)^\ddag = \Xi^2$ for $\Xi^2$ of
\eqref{eq: xi^2 marginal}, which we used above.
Note that
\begin{equation}
e^{i(\Xi^2)^\ddag K}=\frac{1}{K}\,e^{iK(\Xi^2)^\ddag}K
=\frac{1}{K}\left(\sqrt{K}\sqrt{\Xi^1}\frac{1}{\sqrt{K\Xi^1}}\right)K
=\frac{1}{\sqrt{K}}\sqrt{\Xi^1}K\frac{1}{\sqrt{\Xi^1 K}},
\end{equation}
where we have used the double conjugation of the first equation of
\eqref{eq: xi^2 marginal} at the second equality.
Multiplying this and \eqref{eq: xi^2 marginal} together, we obtain
\begin{equation}
e^{-i\Xi^2 K}\,e^{i(\Xi^2)^\ddag K}
=1 .
\label{eq: Xi^2 is real}
\end{equation}
This implies the reality of $\Xi^2$.\footnote{
$\Xi^2$ has the arbitrariness of adding it an integer multiple of
$2\pi/K$
since it appears only in the forms of $e^{\pm i\BRST\xi^2}$,
$e^{iK\Xi^2}$ and $e^{-i\Xi^2 K}$ in $c(\xi)$
in \eqref{eq: representation xi}.
Therefore, \eqref{eq: Xi^2 is real} is the precise expression of the
reality of $\Xi^2$.}

In the above, we adopted \eqref{eq:F=sqrtXi^1} as $F$ satisfying
\eqref{eq:FF^ddag=Xi^1}. We saw that this $F$ satisfies the condition
\eqref{eq: BBRST F} and that the corresponding $\Xi^2$ is real.
However, we can take a more general $F$ of the following form:
\begin{equation}
F=\sqrt{\Xi^1}\,e^{if(\Xi^1)} ,
\label{eq:moregenF}
\end{equation}
where $f(\Xi^1)$ is an arbitrary real function of $\Xi^1$ alone.
This $F$ satisfies \eqref{eq:FF^ddag=Xi^1}, and, as seen by looking
back upon the above proofs of \eqref{eq: BBRST F} and the reality of
$\Xi^2$ for $F=\sqrt{\Xi^1}$,
we can easily see that the two properties hold also for $F$ of
\eqref{eq:moregenF}.
Note, in particular, that the series \eqref{eq: a_n} satisfies
\eqref{eq: BBRST F} for any coefficients $a_n$.
The same $[\Psi_1(K_\tv,B_\tv,c_\tv)]_\xi = \Psi_\mathrm{KOS}$ is
reproduced for any $f(\Xi^1)$.


  Finally, $\xi^{1,2}_\tv$ for generating the KOS
  solution from the single D-brane solution \eqref{eq: D-brane} on the
  tachyon vacuum are obtained from $\Xi^{1,2}(K,V)$ we have determined
  above as $\xi^{1,2}_\tv=U\xi^{1,2} U^{-1}=B_\tv\Xi^{1,2}_\tv$ with
  $\Xi_\tv^{1,2}=\Xi^{1,2}(K_\tv,V_\tv)$.
  Expressed in terms of the original $(K,B,c,V)$, $\xi^{1,2}_\tv$ is given by
\begin{align}
\xi^{1,2}_\tv=UB\Xi^{1,2} U^{-1}=B\sqrt{\frac{K}{1+K}}\Xi^{1,2}\sqrt{\frac{K}{1+K}}.
\end{align}
Note that, although $\Xi^1$ \eqref{eq: xi^1 marginal} is singular at
$K=0$, this singularity is resolved in $\xi^1_\tv$.
This is the case also for $\xi^2_\tv$ obtained from $\Xi^2$ of \eqref{eq: xi^2 marginal}.
The singularities at $K=0$ appearing in various places in our construction are artifacts of solving \eqref{eq: marginal condition} for $\xi^{1,2}$ instead of the original \eqref{eq: original KOS condition} for $\xi_\tv^{1,2}$.

\subsection{Reproducing Erler-Maccaferri solution}
The Erler-Maccaferri solution \cite{Erler:2014eqa} is a generalization of the KOS solution to the case where $V$ is not necessarily regular.
In this case, the boundary condition changing operators are modified to be accompanied by some matter operators, resulting in that $\sigma\bar\sigma = 1$ and the associativity\footnote{For example, $(\sigma\bar\sigma)\sigma \neq \sigma(\bar\sigma\sigma)$ in general.} no longer hold, while $\bar\sigma\sigma = 1$ is kept.
Since \eqref{eq: sigma and V} does not hold anymore, we have to express all equations in terms of $\sigma$ and $\bar\sigma$, not using $V$.
Despite those differences, the solution is still of the same form as \eqref{eq: KOS}:
\begin{align}
	\Psi_\mathrm{EM} = -\frac{1}{\sqrt{1+K}}c(1+K)\sigma \frac{B}{1+K}\bar\sigma(1+K)c\frac{1}{\sqrt{1+K}}.\label{eq: EM}
\end{align}
Our problem is to find $\xi$ satisfying $[\Psi_1(K_\tv,B_\tv,c_\tv)]_\xi = \Psi_\mathrm{EM}$.

Even in the present case, we can reach \eqref{eq: condition on Xi^1} with no obstacle, and hence we obtain\footnote{
According to \cite{Erler:2014eqa}, $\bar\sigma\sigma = 1$ but in general $\sigma\bar\sigma = g \neq 1$ with $g$ being a constant.
Using this fact, we have $S^{-1} = (\sigma(1+K)\bar\sigma)/g$ for $S$ defined in \eqref{eq: SW}.}
\begin{align}
	\Xi^1 =1+ \frac{1}{\sqrt{K}}\l[\l(\sigma\frac{1}{1+K}\bar\sigma \r)^{-1} -1-K\r]\frac{1}{\sqrt{K}}.
	\label{eq: xi^1 EM}
\end{align}
Following the previous process, we get \eqref{eq: xi^2 condition} without any change.
Therefore, we naively expect that $\Xi^2$ is again given by \eqref{eq: xi^2 marginal}.
The only difference from the KOS case is the process of showing that $F$ of \eqref{eq:F=sqrtXi^1} satisfies \eqref{eq: BBRST F}, because we explicitly used $V$ there.

Let us show that \eqref{eq: BBRST F} is satisfied by $F=\sqrt{\Xi^1}$ here as well.
We define
\begin{align}
	S := \sigma \frac{1}{1 + K}	\bar\sigma,\qquad \qquad
	W := S^{-1} - (1+K).\label{eq: SW}
\end{align}
Since $\Xi^1$ is written as
\begin{align}
	\Xi^1 = 1 + \frac{1}{\sqrt{K}}W\frac{1}{\sqrt{K}},
\end{align}
we have only to show
\begin{align}
	\BRST W = 	\cmt{K}{cW},\qquad\quad
	\cmt{c}{W}=0,
	\label{eq: BRST W}
\end{align}
which correspond to \eqref{eq: BRST V} and $\cmt{c}{V}=0$.
The latter relation can be shown as follows:
\begin{align}
	S\cmt{c}{W}S &= Sc - Sc(1+K)S - cS + S(1+K)cS\nonumber\\
	&= \sigma\lcmt{\frac{1}{1+K}}{c}\bar\sigma + \sigma\frac{1}{1+K} \bar \sigma\cmt{1+K}{c}\sigma\frac{1}{1+K}\bar\sigma\nonumber\\
	&= 0.
\end{align}
In the last equality, we have used that $\cmt{1+K}{c} = -\partial c$ is a ghost operator and commutes with $\sigma$.
Next let us consider the former relation in \eqref{eq: BRST W}.
We have
\begin{align}
	\BRST W = -S^{-1}(\BRST S)S^{-1} =-S^{-1}\l(c\cmt{1+K}{\sigma}\frac{1}{1+K}\bar\sigma + \sigma\frac{1}{1+K}\cmt{1+K}{\bar\sigma}c\r)S^{-1},
\end{align}
where we have used $\BRST\sigma = c\cmt{K}{\sigma} =c \cmt{1+K}{\sigma}$ and $\cmt{c}{\cmt{1+K}{\sigma}}=0$ which follow from that $\sigma$ is a matter primary operator of weight 0 ($\bar\sigma$ as well).
As $\sigma$ and $\bar\sigma$ commute with $c$, we obtain the desired relation:
\begin{align}
	\BRST W =  -S^{-1}c(1+K) + (1+K)cS^{-1} = \cmt{K}{cW}+\cmt{c}{W}(1+K) = \cmt{K}{cW}.
\end{align}

As indicated in \cite{Erler:2014eqa}, \eqref{eq: EM} includes a solution which describes the change of the dimension of the D-brane or of the number of D-branes (see also \cite{Kishimoto:2014yea} for the latter).
In the latter case, $\sigma$ and $\bar\sigma$ have multi-components, but our construction here is still valid because $\xi$ can contain arbitrary number of matter operators.
Since \eqref{eq: EM} is expected to describe any backgrounds, the discussion here implies that any open string backgrounds can also be described by \textit{KBc}$_\tv$-representations and \eqref{eq: tachyon Psi_1}, namely, $[\Psi_1(K_\tv,B_\tv,c_\tv)]_\xi$.

\section{Discussions}\label{sec: discussions}
In this paper, we showed that there are various representations of \textit{KBc} algebra which contains matter operators, both on a single D-brane and the tachyon vacuum.
Each of the KOS and the Erler-Maccaferri solutions was reproduced by modifying the single D-brane solution on the tachyon vacuum to the one in a proper representation.
However, though our framework has an ability to determine the representation that reproduces a given solution, there is no guiding principle in our framework to find new physically meaningful solutions.
Here we will discuss two possible directions to address this subject.

The first approach is to find the basic representation of other CFTs.
In SFT, there exists a solution corresponding to any world-sheet CFT.
In addition, given a CFT, there is always the basic representation, and hence $\Psi_0$ \eqref{eq: tachyonj vacuum solution} written by it gives the tachyon vacuum solution of the D-brane system described by the CFT.
Since $\Psi_0$ consists of the basic representation, we can construct any backgrounds if we find the basic representation of each CFT in the language of the reference CFT.
Our method has provided how to find various \textit{KBc} triads, thus has opened a new way to realize this idea.

However, what we have found in this paper are representations on a single D-brane and on the tachyon vacuum, hence our method at glance does not seem to connect \textit{KBc} algebras of different CFTs.
But, there is a clue.
Recalling how we have obtained \textit{KBc}$_\tv$-representations from \textit{KBc}-representations, we can use a unitary transformation to go to representations in other CFTs.
Let us consider a solution of the form $[\Psi_1]_\xi = U(\xi)_\tv^{-1}Q_\tv U(\xi)_\tv $, where $U(\xi)_\tv$ denotes the unitary operator obtained from $U_\tv $ \eqref{eq: tv U} by replacing $(K_\tv ,\cdots)$ with $(K(\xi)_\tv ,\cdots)$.
Following the same logic as in subsection \ref{subsec: tv rep}, a new triad
\begin{align}
	U(\xi)_\tv^{-1}(K(\xi)_\tv, B(\xi)_\tv, c(\xi)_\tv) U(\xi)_\tv ,
\end{align}
forms a representation of \textit{KBc} algebra with its BRST operator given as $Q_\xi := Q_\tv + [[\Psi_1]_\xi,\cdot \}$.
Starting with this triad, we can again generate a class of other representations, among which the basic representation of \textit{KBc} algebra in the CFT corresponding to $[\Psi_1]_\xi$ might exist.
Thus, our method has a potential connection among different CFTs.

The other direction is to investigate the manifold-like structure explained in our previous work \cite{Hata:2021lqz}.
Witten's open SFT has a similar structure to Chern-Simons theory, and several correspondences are known. 
For example, the star product corresponds to the wedge product, and the BRST operator to the exterior derivative.
Though restricted to the \textit{KBc} algebra, the interior product, Lie derivative and Wilson line in SFT were introduced in \cite{Hata:2021lqz}.
The Lie derivative there can generate \textit{KBc}-representations which do not include matter operators, so we defined the \textit{KBc manifold} whose points are those representations.

As explained in appendix \ref{app: KBc construction}, representations introduced in this paper are obtained by extending the previous construction in \cite{Hata:2021lqz}.
In appendix \ref{app: KBc construction}, we redefine the interior product and Lie derivative so that the resultant \textit{KBc}-representations contain matter operators.
However, for incorporating matter operators, we have to ignore some of their properties which were previously respected.
For example, the interior product was nilpotent before, but now its second operation is not defined, because the operation of the interior product on matter operators is not defined (see appendix \ref{app: KBc construction}).\footnote{This fact does not affect our results in this paper.
We discuss the operation of the interior product and the Lie derivative on matter operators in appendix \ref{app: KBc manifold}.}

The \textit{KBc} manifold seems mathematically parallel to the ordinary manifold to some extent, hence it could be possible for it to have a more profound structure.
In order to recover the lost properties, we have to impose further conditions on $\xi$, meaning that representations we obtain are limited.
If the \textit{KBc} manifold is really a physical object, some proper restrictions may lead us to find a correct family of \textit{KBc}-representations, which is expected to have a control over classical solutions in SFT.

\subsection*{Acknowledgement}
D.T.\ thanks Koji Hashimoto for valuable comments on his talk at ``Strings and Fields 2021" in YITP, Kyoto.
The work of D.T. was supported by Grant-in-Aid for JSPS Fellows No.\ 22J20722.
J.Y.\ thanks Hiroshi Kunitomo for daily discussions on SFT.

\appendix
\section{Generating \textit{KBc}-representations}\label{app: KBc construction}
We extend the method in \cite{Hata:2021lqz} to the case where matter operators are included in \textit{KBc}-representations.
We first find an operation lowering the ghost number by $1$, which is an analog of \textit{interior product} in differential geometry.
Second, we define an analog of \textit{Lie derivative} by using the interior product, which provides a differential equation to generate \eqref{eq: representation xi}.

\subsection*{The interior product}
We first construct as general an operation as possible that is linear and lowers the ghost number by $1$, which we write $\interior{X}$.
The subscript $X$ is a (yet undetermined) quantity which characterizes the operation.
Let us start with the following ansatz:
\begin{enumerate}
	\item \label{item: operation on KBc} The operation of $\interior{X}$ on $(K,B,c)$ is written by $K,B,c$ and matter operators.
	\item \label{item: preservation of commutators} Acted by $\interior{X}$, the both hand sides of each relation in \eqref{eq: commutators} return the same result.
	\item \label{item: Leibniz} Against any product of the original $(K,B,c)$, $\interior{X}$ acts according to the (anti-)Leibniz rule just as $\BRST$ does.
\end{enumerate}

One might think it is a problem that we have not defined how $\interior{X}$ acts on matter operators.
But, we will never need such operations.
Here, we do not consider the ``\textit{KBc} manifold" as we did in \cite{Hata:2021lqz}.
Instead, we only focus on finding \textit{KBc}-representations including matter operators.
If one would like to construct the manifold, $\interior{X}$ will be further restricted than what we will obtain here.
The operation of $\interior{X}$ on matter operators is discussed in appendix \ref{app: KBc manifold}.

The generic form of $\interior{X}(K,B,c)$ is given as follows by ansatz \ref{item: operation on KBc}:
\begin{align}
	\interior{X}K = iB f,\qquad\quad
	\interior{X}B = 0,\qquad\quad
	\interior{X}c = g +\sum_{n}h_L^{(n)}Bch_R^{(n)}.\label{eq: pre interior}
\end{align}
Here $f,g$ and $h_{L,R}^{(n)}$ are general operators consisting of $K$ and matter operators.
We determine $f,g$ and $h_{L,R}^{(n)}$ from ansatz \ref{item: preservation of commutators}, by using ansatz \ref{item: Leibniz} for $\interior{X}$ acting on products.
Among the relations in \eqref{eq: commutators}, $\cmt{K}{B} = 0$ and $B^2 = 0$ do not impose any condition on those undetermined quantities; ansatz \ref{item: preservation of commutators} is trivial for the two relations, because there is no quantity carrying ghost number less than $-1$.

Next, let us consider ansatz \ref{item: preservation of commutators} for $c^2 = 0$.
The linearity of $\interior{X}$ gives $\interior{X}0 = 0$, and hence $\interior{X}c^2$ must vanish.
As a necessary condition, $B\interior{X} c^2 = 0$ must also hold, which reads
\begin{align}
	gBc - Bc g- \sum_{n}h_L^{(n)}Bch_R^{(n)}=0.\label{eq: h_12}
\end{align}
From this, $\interior{X}c$ is now written as
\begin{align}
	\interior{X}c = g+ \cmt{g}{Bc} = \acmt{c}{gB}.\label{eq: Ic}
\end{align}
By using the Jacobi identity for (anti-)commutators, one can straightforwardly show $\interior{X}c^2 = 0$ from \eqref{eq: Ic}, meaning that \eqref{eq: h_12} is sufficient.
Under \eqref{eq: Ic}, let us show that $\interior{X}\acmt{B}{c} = \interior{X}1$ holds for any $f$ and $g$.
For the r.h.s., we have $\interior{X}1 = 0$ because of the Leibniz rule.
From \eqref{eq: Ic}, the l.h.s.\ also vanishes as follows:
\begin{align}
	\interior{X}\acmt{B}{c} = \cmt{\acmt{c}{gB}}{B} = gBcB-BcgB = 0.
\end{align}

Note that both $f$ and $g$ are always multiplied by $B$ in $\interior{X}(K,B,c)$.
Thus, defining $X^1:= Bf=fB$ and $X^2 := Bg=gB$, the final result is
\begin{align}
	\interior{X}K = iX^1,\qquad\quad
	\interior{X}B = 0,\qquad\quad
	\interior{X}c = \acmt{c}{X^2},\label{eq: interior}
\end{align}
where $X^{1,2}$ consists of $B$, $K$ and matter operators, and its ghost number is $-1$.\footnote{
$X^1$ and $X^2$ in this paper correspond to $BX^1$ and $B(X^2/K)$ in
\cite{Hata:2021lqz} (see (2.6) there), respectively.
Recalling the correspondence between the form number in differential geometry and the ghost number in open SFT, which was indicated in \cite{Hata:2021lqz}, it is natural that $X^{1,2}$ which we expect is an analog of the tangent vector has ghost number $-1$. 
}

\subsection*{Lie derivative}
The definition of the Lie derivative here is
\begin{align}
	\Lie{X} := -i\acmt{\BRST}{\interior{X}}.\label{eq: def of Lie}
\end{align}
This is an analog of the ordinary Lie derivative given by the anti-commutator between the interior product and the exterior derivative.
Note that $\interior{X}$ and $\Lie{X}$ are defined only for the original $(K,B,c)$ and not for matter operators.
Since the operation of $\interior{X}$ is not closed in the \textit{KBc} algebra while $\BRST$ is, one may be afraid that our definition of $\interior{X}$ and $\Lie{X}$ is incomplete.
However, this is not a problem for our purpose of constructing \textit{KBc}-representations as we shall see below.
Here is the result of the action of $\Lie{X}$ on $(K,B,c$):
\begin{align}
	\Lie{X}K = \BRST X^1,\qquad
	\Lie{X}B = X^1,\qquad
	\Lie{X}c = -c(X^1 + i\cmt{X^2}{K})c + i\cmt{c}{\BRST X^2}.\label{eq: Lie derivative}
\end{align}

The key property of the Lie derivative is that it keeps the \textit{KBc} algebra in the following sense.
Let us consider a new triad,
\begin{align}
	(\tilde K,\tilde B,\tilde c) := (K,B,c) + \epsilon \Lie{X}(K,B,c),\label{eq: tilde}
\end{align}
with $\epsilon$ infinitesimal.
Note that 
\begin{align}
	\cmt{\BRST}{\Lie{X}} = 0\label{eq: BRST Lie}
\end{align}
holds due to $\BRST^2 = 0$.
Using this, we see that \eqref{eq: BRST operations} holds for $(\tilde K,\tilde B,\tilde c)$ to $O(\epsilon)$; for example,
\begin{align}
	\BRST \tilde B = (1+\epsilon \Lie{X})\BRST B = (1 + \epsilon\Lie{X})K = \tilde K.
\end{align}
In addition, one can easily confirm that $\Lie{X}$ follows the Leibniz rule, so \eqref{eq: commutators} is satisfied by the new triad to $O(\epsilon)$; for example,
\begin{align}
	\acmt{\tilde B}{\tilde c} = \acmt{B}{c} + \epsilon \Lie{X}\acmt{B}{c} = 1 + \epsilon\Lie{X}1 = 1.
	\label{eq: tilde Bc}
\end{align}

\subsection*{Generating \textit{KBc}-representations}
What we have learned above is that our Lie derivative can generate new \textit{KBc}-representations which are infinitesimally changed from the basic representation.
Thus we expect that we obtain finitely changed \textit{KBc}-representations if we successively apply the Lie derivative.

However, since we have not defined the action of the Lie derivative on $X$ containing matter operators, such a naive successive operation of Lie derivative is forbidden.
In other words, we have to consider an appropriate Lie derivative on \eqref{eq: tilde} and on the series of subsequent triads.
To construct such a proper new derivative, let us consider the following new interior product:
\begin{align}
	\tinterior{Y}\tilde K = iY^1,\qquad\quad
	\tinterior{Y}\tilde B = 0,\qquad\quad
	\tinterior{Y}\tilde c = \acmt{\tilde c}{Y^2},\label{eq: Lie KBc}
\end{align}
with $Y^{1,2}$ carrying ghost number $-1$.
We again demand that $\tinterior{Y}$ follows the (anti-)Leibniz rule, and define a new Lie derivative,
\begin{align}
  \tLie{Y} := -i\acmt{\BRST}{\tinterior{Y}}.
\end{align}
From this definition and the (anti-)Leibniz rule of $\tinterior{Y}$, we find that $\tLie{Y}$ commutes with $\BRST$ and that $\tLie{Y}$ follows the Leibniz rule, respectively.
This means that 
\begin{align}
	(1+\epsilon \tLie{Y})(\tilde K,\tilde B,\tilde c)\label{eq: tilde tilde}
\end{align}
is also a \textit{KBc}-representation.
The action of $\tLie{Y}$ on \eqref{eq: tilde} is given by
\begin{align}
	\tLie{Y}\tilde K = \BRST Y^1,\qquad
	\tLie{Y}\tilde B = Y^1,\qquad
	\tLie{Y}\tilde c = -\tilde c(Y^1 + i\cmt{Y^2}{\tilde K})\tilde c + i\cmt{\tilde c}{\BRST Y^2},
	\label{eq: tilde Lie}
\end{align}
which is of the same form as \eqref{eq: Lie KBc}.

Thus, we conclude that the triad $(K_s,B_s,c_s)$ with parameter $s$ which is determined through the following differential equations is a \textit{KBc}-representation:
\begin{align}
	&\dot K_s = \BRST \dot \xi^1(s),\qquad
	\dot B_s = \dot\xi^1(s),\nonumber\\
	& \dot c_s = -c_s(\dot \xi^1(s) + i\cmt{\dot\xi^2(s)}{K_s})c_s + i\cmt{c_s}{\BRST\dot\xi^2(s)}.
	\label{eq: differential eq}
\end{align}
Here each of $\xi^{1,2}(s)$ is a function of $s$ which consists of $K,B$ and matter operators, and the dot means $s$-derivative.
The problem left for us is to solve \eqref{eq: differential eq}.
We put $(\xi^1(0),\xi^2(0)) = (B,0)$ and adopt as the initial condition $(K_0,B_0,c_0)=(K,B,c)$.

The first two equations in \eqref{eq: differential eq} can be solved easily to give
\begin{align}
	K_s = \BRST \xi^1(s),\qquad\qquad
	B_s = \xi^1(s).\label{eq: K_sB_s}
\end{align}
Then using
\begin{align}
	\cmt{\dot\xi^2(s)}{K_s} = \cmt{\dot\xi^2(s)}{\BRST\xi^1(s)} = \cmt{\BRST\dot\xi^2(s)}{\xi^1(s)}
	 = \cmt{\BRST\dot\xi^2(s)}{B_s},
\end{align}
we can rewrite the last equation in \eqref{eq: differential eq} as
\begin{align}
	\dot c_s = -c_s(\dot B_s + i\cmt{\BRST\dot\xi^2(s)}{B_s})c_s + i\cmt{c_s}{\BRST\dot\xi^2(s)}.
	\label{eq: simple dot c_s}
\end{align}
Since our construction of $(K_s,B_s,c_s)$ assures that it must be a \textit{KBc}-representation, we can use the \textit{KBc} algebra among them  in advance.
In concrete, using $\acmt{B_s}{c_s} = 1$, we can simplify the differential equation \eqref{eq: simple dot c_s} to
\begin{align}
	\dot c_s = -c_s\dot B_s c_s + i\cmt{c_s(\BRST\dot\xi^2(s))c_s}{B_s}.\label{eq: reduced c_s}
\end{align}

To solve this,  we multiply it by $B_s$ from the left side to obtain
\begin{align}
	\frac{\d }{\d s}(B_sc_s) = -i\cmt{\BRST\dot\xi^2(s)}{B_sc_s}.
\end{align}
Here we have used $B_s\dot \xi^{1,2}(s) = 0$.
This is solved as
\begin{align}
	B_sc_s = E(s)^{-1}Bc E(s),\quad
	E(s) := \mathcal S\exp\l[i\int_0^s\d s'\,\BRST \dot\xi^2(s')\r],\label{eq: B_sc_s}
\end{align}
where  $\mathcal S$ denotes $s$-ordering which puts an operator with larger $s$ to the right.
Using $\acmt{B_s}{c_s} = 1$ again, we also have
\begin{align}
	c_sB_s = E(s)^{-1}cBE(s).\label{eq: c_sB_s}
\end{align}
By expressing $\xi^1(s)$ as $\xi^1(s) = B\Xi^1(s)$ with $\Xi^1(s)$ consisting of $K$ and matter operators (this is always possible), \eqref{eq: B_sc_s} reads
\begin{align}
	B c_s = \frac{1}{\Xi^1(s)} E(s)^{-1}Bc E(s).
\end{align}
Multiplying this by $\eqref{eq: c_sB_s}\times c$ from the left side and using $B_s cB = B_s$ and $c_sB_sc_s = c_s$, we obtain
\begin{align}
	c_s = E(s)^{-1} cB E(s) c \frac{1}{\Xi^1(s)}E(s)^{-1}Bc E(s).\label{eq: c_s}
\end{align}
Since the original differential equation for $c_s$ is a first-order ODE, we conclude that \eqref{eq: c_s} is the unique solution we have been looking for.
Note that $c_s$ of \eqref{eq: c_s} is not determined only by $\xi(s)$ but it in general depends on the whole of $\xi^2(s')$ with $0\le s'\le s$.

If we choose as $\xi(s)$ a special one with
\begin{align}
	\xi^1(1) = \xi^1,\qquad\quad
	\xi^2(s) = s\xi^2\label{eq:choose xi}
\end{align}
for given $\xi^{1,2}$ ($\xi^1(s)$ for $0<s<1$ can be arbitrary), the triad $(K_1,B_1,c_1)$ turns out to be that of \eqref{eq: representation xi}, the representation introduced in the text (we need \eqref{eq: B and xi} to obtain $c(\xi)$ in \eqref{eq: representation xi} from $c_1$).
One can also confirm directly that \eqref{eq: K_sB_s} and \eqref{eq: c_s} form a \textit{KBc}-representation, by following the same procedure shown in the text (subsection \ref{subsec: KBc on D-brane}).


\section{Proof of (\ref{eq: xi^2 condition}) \texorpdfstring{$\Rightarrow$}{TEXT} (\ref{eq: marginal condition})}
\label{app: EM detail}

In section \ref{sec: solutions}, \eqref{eq: xi^2 condition} was derived as a necessary condition for  \eqref{eq: marginal condition} by multiplying it by $B$ from the left side.
Here, we show that \eqref{eq: xi^2 condition} is also sufficient for \eqref{eq: marginal condition}.

First, note that, using $(\Xi^{1,2})^\ddag = \Xi^{1,2}$, the double
conjugation of \eqref{eq: xi^2 condition} reads
\begin{align}
\frac{1}{\sqrt{\BRST\xi^1}}e^{-i\BRST \xi^2}c Be^{iK\Xi^2}\sqrt{K\Xi^1}
=\frac{1}{\sqrt{K}}cB\sqrt{K} .
	\label{eq: xi^2 condition_conj}
\end{align}
Plugging \eqref{eq: representation xi} into the l.h.s.\ of
\eqref{eq: marginal condition} and using \eqref{eq: xi^2 condition}
and \eqref{eq: xi^2 condition_conj},
we obtain
\begin{align}
&\frac{1}{\sqrt{K(\xi)}}c(\xi)\frac{K(\xi)^2}{1 + K(\xi)}B(\xi)c(\xi)\frac{1}{\sqrt{K(\xi)}}\nonumber\\
&=\frac{1}{\sqrt{\BRST\xi^1}}e^{-i\BRST \xi^2}c
e^{iK\Xi^2}\sqrt{K\Xi^1}\sqrt{K}\sigma\frac{1}{1+K}\bar\sigma\sqrt{K}
\left(\sqrt{\Xi^1K}e^{-i\Xi^2K}Bc e^{i\BRST\xi^2}\frac{1}{\sqrt{\BRST \xi^1}} \right)
        \nonumber\\
&=\frac{1}{\sqrt{\BRST\xi^1}}e^{-i\BRST \xi^2}c
e^{iK\Xi^2}\sqrt{K\Xi^1}\sqrt{K}\sigma\frac{1}{1+K}\bar\sigma
KBc\frac{1}{\sqrt{K}}
\nonumber\\
&= \left(\frac{1}{\sqrt{\BRST\xi^1}}e^{-i\BRST \xi^2}cBe^{iK\Xi^2}\sqrt{K\Xi^1} \right)
\sqrt{K}\sigma\frac{1}{1+K}\bar\sigma Kc\frac{1}{\sqrt{K}}
\nonumber\\
&=\frac{1}{\sqrt{K}}cB\sqrt{K}
\sqrt{K}\sigma\frac{1}{1+K}\bar\sigma
Kc\frac{1}{\sqrt{K}}
=\frac{1}{\sqrt{K}}cK\sigma\frac{B}{1+K}\bar\sigma Kc\frac{1}{\sqrt{K}} ,
\end{align}
which is nothing but the r.h.s\ of \eqref{eq: marginal condition}.
Note that we have used \eqref{eq: xi^2 condition} and \eqref{eq: xi^2 condition_conj} at the second and the fifth equalities, respectively, for quantities inside the parentheses.

\section {Interior product and Lie derivative for matter operators}\label{app: KBc manifold}
We extend the action of the interior product and the Lie derivative to matter primary operators.
For simplicity, we consider only one matter operator $V$ with weight $h$.
This $V$ satisfies the following relations:
\begin{align}
    \cmt{B}{V}=0,\qquad \cmt{c}{V}=0, \label{eq: V commutators}\\
    \BRST V=c\cmt{K}{V}+h\cmt{K}{c}V. \label{eq: QV}
\end{align}
We call \eqref{eq: commutators}, \eqref{eq: BRST operations}, \eqref{eq: V commutators} and \eqref{eq: QV} together \textit{KBcV} algebra.

First, we extend the interior product $\interior{X}$ introduced in appendix \ref{app: KBc construction} to define $\interior{X}V$.
In addition to the three ansatz in appendix \ref{app: KBc construction}, we impose the following two:
\begin{enumerate}
    \item[4.]$\interior{X}V$ is written by $K,B,c$ and the matter operator $V$.
    \item[5.] Acted by $\interior{X}$, the both hand sides of each relation in \eqref{eq: V commutators} return the same result.
\end{enumerate}
The generic form of $\interior{X}V$ is given by ansatz 4 as
\begin{align}
    \interior{X}V=BJ,
\end{align}
where $J$ is a general operator written by $K$ and $V$.
We determine $J$ through ansatz 5.
Since $\interior{X}\cmt{B}{V}=0$ automatically holds, we just have to consider $\interior{X}\cmt{c}{V}=0$. From \eqref{eq: interior} and \eqref{eq: interior V}, we have
\begin{align}
    \interior{X}\cmt{c}{V}=\cmt{\acmt{c}{X^2}}{V}-\acmt{c}{BJ}
    =\acmt{c}{\cmt{X^2}{V}-BJ}=0.
\end{align}
Therefore, choosing $BJ=\cmt{X^2}{V}$, $\interior{X}V$ is determined as
\begin{align}
    \interior{X}V=\cmt{X^2}{V}.\label{eq: interior V}
\end{align}
Note that, although $\interior{X}$ respects \eqref{eq: commutators} and \eqref{eq: V commutators}, it does not necessarily keep other relations. 
Especially, the relation $\cmt{V}{\cmt{K}{c}}=0$ which follows from $\cmt{K}{c}=-\partial c$
is not kept by $\interior{X}$, $\interior{X}\cmt{V}{\cmt{K}{c}}\neq 0$.

We can also define the Lie derivative for $V$ as we did for $(K,B,c)$ in appendix \ref{app: KBc construction}:
\begin{align}
    \Lie{X}V&=-i\acmt{\BRST}{\interior{X}}V\nonumber\\
    &=-i\cmt{\BRST X^2}{V}+\acmt{X^1-i\cmt{K}{X^2}}{cV}+(h-1)\acmt{X^1-i\cmt{K}{X^2}}{c}V.\label{eq: lie V}
\end{align}
Then, let us solve the differential equation,
\begin{align}
    \dot{V_s}=-i\cmt{\BRST\dot{\xi}^2(s) }{V_s}+ \acmt{\dot{\xi}^1(s)-i\cmt{K_s}{\dot{\xi}^2(s)}}{c_sV_s}+(h-1)\acmt{\dot{\xi}^1(s)-i\cmt{K_s}{\dot{\xi}^2(s)}}{c_s}V_s,\label{eq: diff V}
\end{align}
together with \eqref{eq: differential eq}, by adopting $V_0 =V$ as the initial condition.
The resulting $(K_s,B_s,c_s,V_s)$ constitute a $KBcV$-representation.

Solving the differential equation \eqref{eq: diff V} for an arbitrary $h$ seems a difficult problem.
However, we can solve it in the special case of $h=1$, where $V$ is a marginal operator. 
Therefore, we restrict the following arguments to $h=1$.
First, we consider 
\begin{align}
    \frac{\d}{\d s}(c_s V_s)=\Lie{\dot{\xi}(s)}^{(s)}(c_s V_s)=-i\acmt{\BRST}{\interior{\dot{\xi}(s)}^{(s)}}(c_s V_s),\label{eq:  diff c_sV_s}
\end{align}
where $\Lie{X}^{(s)}$ and $\interior{X}^{(s)}$ are the Lie derivative and the interior product defined for $(K_s,B_s,c_s,V_s)$.
In particular, $\interior{X}^{(s)}$ is defined by \eqref{eq: interior} and \eqref{eq: interior V} with $(K,B,c,V)$ replaced by $(K_s,B_s,c_s,V_s)$.
Since $\BRST(c_s V_s)=0$ for $h=1$, the differential equation \eqref{eq:  diff c_sV_s} is reduced to
\begin{align}
        \frac{\d}{\d s}(c_s V_s)=-i\BRST\interior{\dot{\xi}(s)}^{(s)}(c_sV_s)=-i[\BRST\dot{\xi}^2(s),c_s V_s]\label{eq: diff eq of cV}.
\end{align}
The solution to \eqref{eq: diff eq of cV} is given by
\begin{align}
    c_s V_s= E(s)^{-1}cVE(s),\label{eq: c_sV_s}
\end{align}
where $E(s)$ is defined in \eqref{eq: B_sc_s}.
Using this result, $V_s$ itself is obtained as follows:
\begin{align}
    V_s =\acmt{B_s}{c_s V_s}=\acmt{\xi^1(s)}{c_s V_s} =\acmt{\xi^1(s)}{E(s)^{-1}cVE(s)}.\label{eq: V_s}
\end{align}
If we choose $\xi(s)$ of \eqref{eq:choose xi}, $V(\xi):=V_1$ is given by
\begin{align}
    V(\xi)=\acmt{\xi^1}{e^{-i\BRST \xi^2}cVe^{i\BRST \xi^2}}.
\end{align}
This $V(\xi)$ together with $(K(\xi),B(\xi),c(\xi))$ of \eqref{eq: representation xi} constitutes to a \textit{KBcV}-representation. 
Starting with a solution containing matter operators, we can construct new solutions by using this representation.

However, $\interior{X}$ and $\Lie{X}$ for matter operators have several problems due to the fact\linebreak $\interior{X}\cmt{V}{\cmt{K}{c}}\neq 0$.
The first problem is that $\Lie{X}$ does not necessarily keep $V$ real.
For a real $X$, we have
\begin{align}
    (\Lie{X}(K,B,c))^\ddag=\Lie{X}(K,B,c),
\end{align}
and this property ensures that the triad $(K(\xi),B(\xi),c(\xi))$ is real as long as $\xi$ is real.
On the other hand, for a real $V$, we have from \eqref{eq: lie V}
\begin{align}
    (\Lie{X}V)^\ddag =\Lie{X}V + (h-1)\interior{X}\cmt{V}{\cmt{K}{c}}.
\end{align}
This implies that $V_s$ determined by \eqref{eq: diff V} is no longer real unless $h=1$.

Though this reality problem of $V_s$ is resolved for $V$ with weight $1$, there still remains another problem existent even in the case $h=1$.
This is a problem that the expression of $\Lie{X}V$ is not unique.
This expression depends on the order of $\cmt{K}{c}$ and $V$ in \eqref{eq: QV}.
If we rewrite \eqref{eq: QV} as
\begin{align}
    \BRST V =c\cmt{K}{V}+hV\cmt{K}{c},
\end{align}
 the corresponding $\Lie{X}V=-i\acmt{\BRST}{\interior{X}}V$ is a different quantity from \eqref{eq: lie V}, but is given by
\begin{align}
    \Lie{X}V=-i\cmt{\BRST X^2}{V}-c\cmt{X^1-i\cmt{K}{X^2}}{V}+hV\acmt{X^1-i\cmt{K}{X^2}}{c}.
\end{align}
Of course, if we choose one expression of $\Lie{X}V$, we obtain a family
of \textit{KBcV}-representations.
However, the ordering problem and the resulting ambiguity of $\Lie{X}V$ may cause inconveniences, if one applies the algebraic tools we have developed here to other problems in SFT.

\bibliographystyle{jhep} 
\bibliography{references.bib}

\providecommand{\href}[2]{#2}\begingroup\raggedright\begin{thebibliography}{10}

\bibitem{Witten:1985cc}
E.~Witten, \emph{{Noncommutative Geometry and String Field Theory}},
  \href{https://doi.org/10.1016/0550-3213(86)90155-0}{\emph{Nucl. Phys. B}
  {\bfseries 268} (1986) 253}.

\bibitem{Sen:1998sm}
A.~Sen, \emph{{Tachyon condensation on the brane anti-brane system}},
  \href{https://doi.org/10.1088/1126-6708/1998/08/012}{\emph{JHEP} {\bfseries
  08} (1998) 012} [\href{https://arxiv.org/abs/hep-th/9805170}{{\ttfamily
  hep-th/9805170}}].

\bibitem{Sen:1999mg}
A.~Sen, \emph{{NonBPS states and Branes in string theory}},  in \emph{{Advanced
  School on Supersymmetry in the Theories of Fields, Strings and Branes}},
  pp.~187--234, 1, 1999 [\href{https://arxiv.org/abs/hep-th/9904207}{{\ttfamily
  hep-th/9904207}}].

\bibitem{Schnabl:2005gv}
M.~Schnabl, \emph{{Analytic solution for tachyon condensation in open string
  field theory}}, \href{https://doi.org/10.4310/ATMP.2006.v10.n4.a1}{\emph{Adv.
  Theor. Math. Phys.} {\bfseries 10} (2006) 433}
  [\href{https://arxiv.org/abs/hep-th/0511286}{{\ttfamily hep-th/0511286}}].

\bibitem{Erler:2009uj}
T.~Erler and M.~Schnabl, \emph{{A Simple Analytic Solution for Tachyon
  Condensation}},
  \href{https://doi.org/10.1088/1126-6708/2009/10/066}{\emph{JHEP} {\bfseries
  10} (2009) 066} [\href{https://arxiv.org/abs/0906.0979}{{\ttfamily
  0906.0979}}].

\bibitem{Okawa:2006vm}
Y.~Okawa, \emph{{Comments on Schnabl's analytic solution for tachyon
  condensation in Witten's open string field theory}},
  \href{https://doi.org/10.1088/1126-6708/2006/04/055}{\emph{JHEP} {\bfseries
  04} (2006) 055} [\href{https://arxiv.org/abs/hep-th/0603159}{{\ttfamily
  hep-th/0603159}}].

\bibitem{Kiermaier:2007ba}
M.~Kiermaier, Y.~Okawa, L.~Rastelli and B.~Zwiebach, \emph{{Analytic solutions
  for marginal deformations in open string field theory}},
  \href{https://doi.org/10.1088/1126-6708/2008/01/028}{\emph{JHEP} {\bfseries
  01} (2008) 028} [\href{https://arxiv.org/abs/hep-th/0701249}{{\ttfamily
  hep-th/0701249}}].

\bibitem{Schnabl:2007az}
M.~Schnabl, \emph{{Comments on marginal deformations in open string field
  theory}}, \href{https://doi.org/10.1016/j.physletb.2007.08.023}{\emph{Phys.
  Lett. B} {\bfseries 654} (2007) 194}
  [\href{https://arxiv.org/abs/hep-th/0701248}{{\ttfamily hep-th/0701248}}].

\bibitem{Fuchs:2007yy}
E.~Fuchs, M.~Kroyter and R.~Potting, \emph{{Marginal deformations in string
  field theory}},
  \href{https://doi.org/10.1088/1126-6708/2007/09/101}{\emph{JHEP} {\bfseries
  09} (2007) 101} [\href{https://arxiv.org/abs/0704.2222}{{\ttfamily
  0704.2222}}].

\bibitem{Kiermaier:2007vu}
M.~Kiermaier and Y.~Okawa, \emph{{Exact marginality in open string field
  theory: A General framework}},
  \href{https://doi.org/10.1088/1126-6708/2009/11/041}{\emph{JHEP} {\bfseries
  11} (2009) 041} [\href{https://arxiv.org/abs/0707.4472}{{\ttfamily
  0707.4472}}].

\bibitem{Erler:2007rh}
T.~Erler, \emph{{Marginal Solutions for the Superstring}},
  \href{https://doi.org/10.1088/1126-6708/2007/07/050}{\emph{JHEP} {\bfseries
  07} (2007) 050} [\href{https://arxiv.org/abs/0704.0930}{{\ttfamily
  0704.0930}}].

\bibitem{Okawa:2007ri}
Y.~Okawa, \emph{{Analytic solutions for marginal deformations in open
  superstring field theory}},
  \href{https://doi.org/10.1088/1126-6708/2007/09/084}{\emph{JHEP} {\bfseries
  09} (2007) 084} [\href{https://arxiv.org/abs/0704.0936}{{\ttfamily
  0704.0936}}].

\bibitem{Okawa:2007it}
Y.~Okawa, \emph{{Real analytic solutions for marginal deformations in open
  superstring field theory}},
  \href{https://doi.org/10.1088/1126-6708/2007/09/082}{\emph{JHEP} {\bfseries
  09} (2007) 082} [\href{https://arxiv.org/abs/0704.3612}{{\ttfamily
  0704.3612}}].

\bibitem{Fuchs:2007gw}
E.~Fuchs and M.~Kroyter, \emph{{Marginal deformation for the photon in
  superstring field theory}},
  \href{https://doi.org/10.1088/1126-6708/2007/11/005}{\emph{JHEP} {\bfseries
  11} (2007) 005} [\href{https://arxiv.org/abs/0706.0717}{{\ttfamily
  0706.0717}}].

\bibitem{Kiermaier:2007ki}
M.~Kiermaier and Y.~Okawa, \emph{{General marginal deformations in open
  superstring field theory}},
  \href{https://doi.org/10.1088/1126-6708/2009/11/042}{\emph{JHEP} {\bfseries
  11} (2009) 042} [\href{https://arxiv.org/abs/0708.3394}{{\ttfamily
  0708.3394}}].

\bibitem{Kiermaier:2010cf}
M.~Kiermaier, Y.~Okawa and P.~Soler, \emph{{Solutions from boundary condition
  changing operators in open string field theory}},
  \href{https://doi.org/10.1007/JHEP03(2011)122}{\emph{JHEP} {\bfseries 03}
  (2011) 122} [\href{https://arxiv.org/abs/1009.6185}{{\ttfamily 1009.6185}}].

\bibitem{Erler:2014eqa}
T.~Erler and C.~Maccaferri, \emph{{String Field Theory Solution for Any Open
  String Background}},
  \href{https://doi.org/10.1007/JHEP10(2014)029}{\emph{JHEP} {\bfseries 10}
  (2014) 029} [\href{https://arxiv.org/abs/1406.3021}{{\ttfamily 1406.3021}}].

\bibitem{Erler:2019fye}
T.~Erler and C.~Maccaferri, \emph{{String field theory solution for any open
  string background. Part II}},
  \href{https://doi.org/10.1007/JHEP01(2020)021}{\emph{JHEP} {\bfseries 01}
  (2020) 021} [\href{https://arxiv.org/abs/1909.11675}{{\ttfamily
  1909.11675}}].

\bibitem{Hata:2021lqz}
H.~Hata and D.~Takeda, \emph{{Interior product, Lie derivative and Wilson line
  in the KBc subsector of open string field theory}},
  \href{https://doi.org/10.1007/JHEP07(2021)117}{\emph{JHEP} {\bfseries 07}
  (2021) 117} [\href{https://arxiv.org/abs/2103.10597}{{\ttfamily
  2103.10597}}].

\bibitem{Erler:2010zza}
T.~Erler, \emph{{A simple analytic solution for tachyon condensation}},
  \href{https://doi.org/10.1007/s11232-010-0053-z}{\emph{Theor. Math. Phys.}
  {\bfseries 163} (2010) 705}.

\bibitem{Masuda:2012kt}
T.~Masuda, T.~Noumi and D.~Takahashi, \emph{{Constraints on a class of
  classical solutions in open string field theory}},
  \href{https://doi.org/10.1007/JHEP10(2012)113}{\emph{JHEP} {\bfseries 10}
  (2012) 113} [\href{https://arxiv.org/abs/1207.6220}{{\ttfamily 1207.6220}}].

\bibitem{Erler:2012dz}
T.~Erler, \emph{{The Identity String Field and the Sliver Frame Level
  Expansion}}, \href{https://doi.org/10.1007/JHEP11(2012)150}{\emph{JHEP}
  {\bfseries 11} (2012) 150} [\href{https://arxiv.org/abs/1208.6287}{{\ttfamily
  1208.6287}}].

\bibitem{Hata:2011ke}
H.~Hata and T.~Kojita, \emph{{Winding Number in String Field Theory}},
  \href{https://doi.org/10.1007/JHEP01(2012)088}{\emph{JHEP} {\bfseries 01}
  (2012) 088} [\href{https://arxiv.org/abs/1111.2389}{{\ttfamily 1111.2389}}].

\bibitem{Mertes:2016vos}
N.~Mertes and M.~Schnabl, \emph{{String field representation of the Virasoro
  algebra}}, \href{https://doi.org/10.1007/JHEP12(2016)151}{\emph{JHEP}
  {\bfseries 12} (2016) 151}
  [\href{https://arxiv.org/abs/1610.00968}{{\ttfamily 1610.00968}}].

\bibitem{Kishimoto:2014yea}
I.~Kishimoto, T.~Masuda, T.~Takahashi and S.~Takemoto, \emph{{Open String
  Fields as Matrices}}, \href{https://doi.org/10.1093/ptep/ptv023}{\emph{PTEP}
  {\bfseries 2015} (2015) 033B05}
  [\href{https://arxiv.org/abs/1412.4855}{{\ttfamily 1412.4855}}].

\end{thebibliography}\endgroup
\end{document}